\documentclass[intlimits,twoside,a4paper]{article}

\usepackage[cp1251]{inputenc}

\usepackage[eqsecnum]{cmpj3}

\usepackage{bm}
\usepackage{subfigure}

\issue{2018}{21}{3}{33601}
\doinumber{10.5488/CMP.21.33601}

\title[On the spatially periodic ordering]{On the spatially periodic ordering in the system of electrons above the
surface of liquid helium in an external electric field}
\author[D.M. Lytvynenko, Yu.V. Slyusarenko, A.I. Kirdin]{D.M. Lytvynenko\refaddr{label1,label2},
Yu.V. Slyusarenko\refaddr{label1,label2}, A.I. Kirdin\refaddr{label1}}
\addresses{
\addr{label1}
National Science Center Kharkiv Institute of Physics and Technology,
1 Akademichna St., 61108 Kharkiv, Ukraine
\addr{label2}
Kharkiv V.N. Karazin National University, 4 Svobody Sq., 61022 Kharkiv, Ukraine
}

\date{Received May 14, 2018, in final form August 9, 2018}

\begin{document}

\maketitle

\begin{abstract}
A theory of equilibrium states of electrons above a liquid helium surface in the presence of an external clamping field is built based on the first principles of quantum statistics for the system of many identical Fermi-particles. The approach is based on the variation principle modified for the considered system and on Thomas-Fermi model. In terms of the developed theory we obtain the self-consistency equations that connect the parameters of the system description, i.e., the potential of a static electric field, the distribution function of electrons and the surface profile of a liquid dielectric. The equations are used to study the phase transition of the system to a spatially periodic state. To demonstrate the capabilities of the proposed method, the characteristics of the phase transition of the system to a spatially periodic state of a trough type are analyzed.

\pacs 64.60.Ai, 73.20.At
\end{abstract}

\section{Introduction}

In spite of more than forty year history~\cite{pla1971CrandallW,jetpl1974Shikin,jetp1975MonarkhaS,prl1979FisherHP}, the studies of phenomena associated with the formation of spatially periodic states in a system of charged particles above dielectric surface are still relevant. The possibility of spatially periodic ordering in 3D system of charges (electrons in metals) was predicted by Wigner~\cite{pr1934Wigner}. Due to this research, a new term called ``Wigner crystallization'' (WC) appeared. The other way of reaching the phase transition with the formation of 3D spatially periodic structures in similar systems was shown in \cite{ltp1998PPS}. Unlike 3D systems, the stable spatially periodic states in the system of electrons near the boundary between two media were experimentally registered in \cite{prl1979GrimesA,pla1979LeidererW,ss1982LeidererES,pla1980EbnerL,jetpl1979TroyanovskiiVH,ss1984Kajita,prb1982BishopDT}. The chronology of the research is given in monographs~\cite{book1997Andrei,book2003MonarkhaK} and in review papers~\cite{ltp1982MonarkhaS,ltp2012MonarkhaS,ufn2011Shikin}.

The theoretical papers studying both 2D WC effects~\cite{prl1979GrimesA} and the formation of macroscopic dimple lattices \cite{pla1979LeidererW} are usually based on the concept of the energy spectrum of a single (i.e., ``levitating'') electron above the dielectric surface. This concept considers a single electron above a planar dielectric surface together with its electrostatic image as an analogue of a hydrogen-like atom~\cite{prl1969ColeC}. Obviously, the description of a many-particle system of charges above the dielectric surface using this approach faces mathematical and methodological issues, because the image method is a mathematical technique to avoid a consistent solution of the Poisson equation for a single charge above the metallic or dielectric surface. The mentioned issues can be avoided by describing the system in terms of a microscopic theory. Such a theory must consider a quantum-mechanical system of many particles~\cite{jmp2012LytvynenkoSK,past2012SlyusarenkoSK,cmp2009LytvynenkoS,jps2015SlyusarenkoL,jpa2017LytvynenkoS}, and take into account the external pressing electric field. This field is important in forming such systems, since the field induced by the charges (electrons) in a dielectric is insufficient to keep them near its surface.

The basics of the microscopic approach were formulated in~\cite{jmp2012LytvynenkoSK}. This approach uses a variation principle and the modified Thomas-Fermi model. The approach allows one to obtain  self-consistency equations relating the parameters of such a system description (the potential of a static electric field, the distribution function of charges (electrons) and the profile of a liquid dielectric surface). The comparative results of the developed theory versus the experimental data~\cite{pla1979LeidererW} are in qualitative agreement.

The present paper studies a system of electrons above the liquid helium surface in an external pressing field in terms of the quasi-classical approach presented in~\cite{jmp2012LytvynenkoSK} and developed in~\cite{jps2015SlyusarenkoL,jpa2017LytvynenkoS}. Unlike these papers, the current paper analysis is not limited to non-degeneracy or quasi-neutrality of the system. Quasi-neutrality of the system assumes that the external field is compensated by the field of electrons far from the dielectric surface. This paper considers the ``charged'' problem, where the pressing field can be larger than the electrons are capable of compensating. Unlike papers~\cite{jmp2012LytvynenkoSK,past2012SlyusarenkoSK,cmp2009LytvynenkoS,jps2015SlyusarenkoL}, this research also goes beyond Boltzmann's statistics. The statistical approach to the description of spatially inhomogeneous states in Coulomb systems was also used in papers~\cite{pre1998LevZ,pre2011LevZ,ujp2015LevOTZ,epjb2014LevOTZ}. However, these studies were based on the usage of the modified electrostatic potential of a single electron and on the methods of functional integration to calculate grand statistical sums.

\section{Self-consistency equations for the system of electrons above liquid helium surface}

The present paper research is based on the variation principle proposed in \cite{jmp2012LytvynenkoSK}, so let us remind the main points of the theory. We consider the system of electrons 
with charge $-e$ ($e>0$), mass $m$, spin~${1 / 2}$, momentum ${\bf{p}}$ and energy ${\varepsilon _{\bf{p}}} = {{{{\bf{p}}^2}} / {2m}}$. The electrons are located in vacuum $z > \xi ( {\boldsymbol{\rho}} )$ (region~``1'') above the surface of liquid helium film $\xi ( {\bf{\rho }} ) > z >  - d$ (region~``2'') having thickness $d$, dielectric constant $\varepsilon$ and surface tension $\alpha$. Let us assume that the film is located on a flat solid dielectric substrate (region~``3'') with dielectric constant ${\varepsilon _d} \gg \varepsilon$. The surface profile of the helium film is described by function $\xi ( {\boldsymbol{\rho} } ) \equiv \xi ( {x,y} )$, where ${\boldsymbol{\rho}} \equiv \{ {x,y} \}$ is the radius vector in $z = 0$ plane of the Cartesian coordinate system $\{ z,x,y\}$. The boundaries between regions ``1''--``3'' along the direction of  ${\boldsymbol{\rho}}$ are  assumed to be unlimited. To avoid questions on the ``repulsion'' of electrons along ${\boldsymbol{\rho}}$, we assume that the system is located in a vessel with walls at $\rho  \to \infty$, forbidding electrons from leaving the system in ${\boldsymbol{\rho}}$ direction.
 An external clamping electric field ${E^\text{(e)}}$ directed along $z$-axis prevents electrons from leaving the system in $z$-direction.
In region ``1'', the system is described by the distribution function ${f_{\bf{p}}}( {\bf{r}} )$ of electrons, their electric potential $\varphi _1^{\text{(i)}}( {\bf{r}} )$, the potential $\varphi _1^\text{(e)}( {\bf{r}} )$ of the clamping field ${E^\text{(e)}}$ and helium surface profile $\xi ({\boldsymbol{\rho }})$. Region~``2'' is described by helium surface profile $\xi ({\boldsymbol{\rho }})$ and the total potential $\varphi _2^{} = \varphi _2^{\text{(i)}} + \varphi _2^\text{(e)}$. Region~``3'' is described by the total potential $\varphi _3^{} = \varphi _3^{\text{(i)}} + \varphi _3^\text{(e)}$ in the solid substrate.

To obtain  self-consistency equations for  equilibrium values of the main parameters ${f_{\bf{p}}}( {\bf{r}} )$, $\xi ({\boldsymbol{\rho }})$ and $\varphi _1^{\text{(i)}}( {\bf{r}} )$ describing the system, it is necessary to obtain the maximum of the system entropy $S$
\begin{equation}
\label{1a6}
S =  - \frac{{{2}}}{{{{\left( {2\piup \hbar } \right)}^3}}}\int {\rd^3{\bf{r}}\,\rd^3{\bf{p}}} \left[ \bar f\ln \bar f +  \left( 1 - \bar f \,\right)\ln \left( 1 - \bar f \,\right) \right],
\qquad
\bar f = \frac{{{{\left( {2\piup \hbar } \right)}^3}}}{2}{f_{\bf{p}}}\left( {\bf{r}} \right),
\end{equation}
if the following conditions take place. First, for a fixed ${E^\text{(e)}}$ value, the total number of electrons $N = \int {\rd^3{\bf{r}}\,\rd^3} {\bf{p}}{f_{\bf{p}}}( {\bf{r}} )$, their total momentum ${\bf{P}} = \int {\rd^3{\bf{r}}\,\rd^3} {\bf{p}}{f_{\bf{p}}}( {\bf{r}} ){\bf{p}}$ (equal to zero, as the system stays at rest) and the total energy of the system~\cite{jmp2012LytvynenkoSK}
\begin{equation}\label{2a6}
{E_\text{t}} = \int\limits_{{V_1}}^{} {\rd^3{\bf{r}}} \left\{ {K -en\left[ {\frac{{\varphi _1^{\text{(i)}}}}{2} + \varphi _1^\text{(e)}} \right] + \frac{{E{{_1^\text{(e)}}^2}}}{{8\piup }}} \right\} + \int\limits_{{V_2}}^{} {\rd^3{\bf{r}}} \frac{{\varepsilon E_2^2}}{{8\piup }} + \int\limits_{{V_3}}^{} {\rd^3{\bf{r}}} \frac{{\varepsilon _d^{}E_3^2}}{{8\piup }} + \frac{\alpha }{2}\int {\rd S\left[ {{{\left( {\nabla \xi } \right)}^2} + {{\left( {\kappa \xi } \right)}^2}} \right]}
\end{equation}
 remain unchanged.
 In~\eqref{2a6} ${K = \int {\rd^3{\bf{p}}{f_{\bf{p}}}{\varepsilon _{\bf{p}}} } }$, ${\bf{E}}_j^{} =  - \nabla \varphi _j^{}$, ${\bf{E}}_j^\text{(e)} =  - \nabla \varphi _j^\text{(e)}$, ${V_j},\;j = 1,2,3,$ are the volumes of regions ``1'', ``2'' and ``3'', respectively, and   $\rd S = {\rd^2}\boldsymbol\rho \{ {1 + {{[ {{\nabla _{\boldsymbol{\rho}}}\xi ( {\boldsymbol{\rho}} )} ]}^2}}\}^{1/2} $, ${\nabla _{\boldsymbol{\rho}}} \equiv {\partial  / {\partial {\boldsymbol{\rho}}}}$, ${\varphi _j} = \varphi _j^{\text{(i)}} + \varphi _j^\text{(e)}$.
 Secondly, in the absence of electrons above the film, its surface profile cannot be transformed. Thirdly, Poisson equation must take place in all three regions of the system.  The electron density in~\eqref{2a6} has the form:
  \begin{equation}
\label{3a6}
n\left( {\bf{r}} \right) = \int {\rd^3{\bf{p}}} {f_{\bf{p}}}\left( {\bf{r}} \right).
\end{equation}
Let us also make the following remark. As the helium film consists of an incompressible fluid, its total volume must be fixed.
However, the interaction of liquid helium film with the electrons pressed by the external electric field to its surface leads to the lowering of the helium surface profile ${\xi ( {\boldsymbol{\rho}} )}$ (see~\cite{pla1979LeidererW,ss1982LeidererES}). Moreover, the surface profile of this deflection remains flat up to a certain critical value of the external clamping field. At a glance, this fact contradicts the incompressibility condition for the volume of liquid helium film. 
In real experiments, the  external electric field is provided by placing a positively charged plate of a flat capacitor into helium below its surface. The linear dimensions of helium surface subsidence are comparable to
the linear dimensions of the plate $L$~\cite{pla1979LeidererW,ss1982LeidererES}. Thus, the decrease of helium volume above the condenser plate
due to the subsidence of the film surface under the action of electrons pressed by the external electric field
must be compensated by its increase outside the plate. In the present paper, we consider the flat helium surface region assuming that the edge effect of an increase of the helium surface profile takes place near the vessel walls at $| {\boldsymbol{\rho}} | \to \infty$. This edge effect being ignored allows one to skip the condition for helium volume incompressibility while solving the variation problem.

The problem of obtaining the conditional maximum of entropy can be replaced by the problem of obtaining  the unconditional minimum of a grand thermodynamic potential $\tilde \Omega$ (for more details, see~\cite{jmp2012LytvynenkoSK}):
\begin{equation}
\label{4a6}
\tilde \Omega  =  - S + {Y_0}E + {Y_i}{P_i} + {Y_4}N + \int {\rd^2{\boldsymbol{\rho}}}\, {\lambda _\xi }\left( {\boldsymbol{\rho}} \right){ {\xi \left( {\boldsymbol{\rho}} \right)} \big|_{N = 0}} + \int {\rd^3{\bf{r}}}\, \lambda \left( {\bf{r}} \right)\left[ {\Delta \varphi \left( {\bf{r}} \right) + 4\piup Qn\left( {\bf{r}} \right)} \right],
\end{equation}
where ${Y_0}$, ${Y_i}$, ${Y_4}$, $\lambda ( {\bf{r}} )$, ${\lambda _\xi }( {\boldsymbol{\rho}} )$ are the Lagrange multipliers corresponding to the above conditions. To obtain the minimum of $\tilde \Omega$, it is necessary to calculate the following conditions for the variation derivatives:
\begin{equation*}
\label{4aa6}
{\left. {\frac{{\delta \tilde \Omega }}{{\delta f}}} \right|_{\xi ,{\varphi ^{\text{(i)}}}}} = 0,
\qquad
{\left. {\frac{{\delta \tilde \Omega }}{{\delta {\varphi ^{\text{(i)}}}}}} \right|_{f,\xi }} = 0,
\qquad
{\left. {\frac{{\delta \tilde \Omega }}{{\delta \xi }}} \right|_{f,{\varphi ^{\text{(i)}}}}} = 0.
\end{equation*}
Solving the variation problem results in the following equations (for details, see~\cite{jmp2012LytvynenkoSK}):
\begin{equation}
\label{5a6}
{\left\{ {\frac{{{2}T}}{\alpha }\int {\rd^3{\bf{p}}\frac{{\ln\left( {1 - \bar f} \,\right)}}{{{{\left( {2\piup \hbar } \right)}^3}}}}  + \frac{\varepsilon }{{8\piup }}\left[ {E{{_2^\text{(e)}}^2} - E_2^2} \right]} \right\}_{z = \xi }} = {\kappa ^2}\xi \sqrt {1 + {{\left( {\nabla \xi } \right)}^2}}  - \nabla \left\{ {\frac{{\nabla \xi \left[ {2 + {\kappa ^2}{\xi ^2} + 3{{\left( {\nabla \xi } \right)}^2}} \right]}}{{2\sqrt {1 + {{\left( {\nabla \xi } \right)}^2}} }}} \right\},
\end{equation}
where the distribution function ${f_{\bf{p}}}( {\bf{r}} )$ is expressed by
 \begin{equation}
\label{6a6}
{f_{\bf{p}}}\left( {\bf{r}} \right) = \theta \big( {z - \xi \left( {\boldsymbol{\rho}} \right)} \big)\frac{{{2}}}{{{{\left( {2\piup \hbar } \right)}^3}}}{\left[ {1 + \exp {T^{ - 1}}\left( {{\varepsilon _{\bf{p}}} - \mu  -e{\varphi _1}} \right)} \right]^{ - 1}},
\vspace{-1mm}
\end{equation}
$\theta ( z )$ is the Heaviside unit function, and $\kappa$ in~\eqref{5a6} is given by~\cite{book1989ShikinM}
 \begin{equation}
\label{7a6}
\kappa \left( d \right) = \sqrt {\frac{\rho }{\alpha }\left( {g + f} \right)},
\qquad
{f = \frac{{{g_0}{d_v}}}{{{d^4}\left( {d + {d_v}} \right)}}\left( {3 + \frac{d}{{d + {d_v}}}} \right)},
\end{equation}
where ${d_v} = 1.65 \cdot {10^{ - 5}}$~cm and ${g_0} = 2.2 \cdot {10^{ - 14}}$~cm$^{5} \cdot \,$s$^{ - 2}$, $g$ is the gravity acceleration, $\alpha$ is the surface tension of the liquid helium, $\rho$ is its density and $f$ is Van der Waals constant, which in the case of a massive liquid helium film ($d \to \infty$) can be neglected compared to $g$. In the case of a thin helium film, the gravitational force acting on helium atoms is negligibly small compared to Van der Waals forces. This situation takes place for films thinner than $d \sim {10^{ - 4}}$~cm.

Equations~\eqref{5a6},~\eqref{6a6} together with the equations for the potentials of the electric fields, both external $\varphi^\text{(e)}( {\bf{r}} )$ and total $\varphi ( {\bf{r}} )=\varphi^{\text{(i)}}( {\bf{r}} )+\varphi^\text{(e)}( {\bf{r}})$
in all three regions of the system:
\begin{equation}
\label{8a6}
\Delta \varphi _1
\left( {\bf{r}} \right) -4\piup en\left( {\bf{r}} \right) = 0,
\quad
\Delta \varphi _2
\left( {\bf{r}} \right) = 0,
\quad
\Delta \varphi _3\left( {\bf{r}} \right) = 0,
\quad
\Delta \varphi _j^\text{(e)}\left( {\bf{r}} \right) = 0,
\quad
j=1,2,3,
\end{equation}
form a system of self-consistency equations. However, it must be supplemented by the boundary conditions for the potentials and electric fields at boundaries $z = \xi ( {\boldsymbol{\rho }} )$ and $z =  - d$. 
Since there are no surface charges at the boundaries, these boundary conditions together with the finiteness conditions for the electric fields have the form:
\begin{equation}
\begin{gathered}
\label{12a6}
{\varphi _1}\left( {z,{\boldsymbol{\rho}}} \right)\left| {_{z = \xi }} \right. = {\varphi _2}\left( {z,{\boldsymbol{\rho}}} \right)\left| {_{z = \xi }} \right.,
\qquad
{\left\{ {\big( {{\bf{n}}({\boldsymbol{\rho}}) \cdot \nabla } \big)\left[ {\varepsilon {\varphi _2}\left( {z,{\boldsymbol{\rho}}} \right) - {\varphi _1}\left( {z,{\boldsymbol{\rho}}} \right)} \right]} \right\}_{z = \xi }} = 0,
\\
{\varphi _2}\left( {z,{\boldsymbol{\rho}}} \right)\big| {_{z =  - d}}  = {\varphi _3}\left( {z,{\boldsymbol{\rho}}} \right)\big| {_{z =  - d}} \,,
\qquad
{\left[ {\varepsilon \frac{{\partial {\varphi _2}\left( {z,{\boldsymbol{\rho}}} \right)}}{{\partial z}} - {\varepsilon _d}\frac{{\partial {\varphi _3}\left( {z,{\boldsymbol{\rho}}} \right)}}{{\partial z}}} \right]_{z =  - d}} = 0,
\qquad
\left| {\frac{{\partial {\varphi _j}}}{{\partial z}}} \right| <  + \infty,
\end{gathered}
\end{equation}
where ${\bf{n}}( {\boldsymbol{\rho}} ) = {[ {1 + {{( {\nabla \xi } )}^2}} ]^{ - {1 / 2}}}\{ { - {{\partial \xi } / {\partial x}}, - {{\partial \xi } / {\partial y}},1} \}$ is the normal to the surface profile $\xi ( {\boldsymbol{\rho}} )$ at point ${\boldsymbol{\rho}}$.
Let us also note that the conditions similar to equation~\eqref{12a6} take place for ${\tilde \varphi _j^\text{(e)}( {z,{\boldsymbol{\rho}}} )}$ ($j=1,2,3$).

\section{Scenario of the phase transition with the formation of spatially periodic structures}

The external electric field $E_{}^\text{(e)}$ that presses the electrons to the helium surface can lead to its subsidence 
in the region of this field action (see, e.g.,~\cite{book1997Andrei,book2003MonarkhaK,ltp1982MonarkhaS,ltp2012MonarkhaS,ufn2011Shikin,jmp2012LytvynenkoSK}). Since the bottom of this deflection remains flat, 
it can be described by  $\bar \xi$ parameter (the subsidence depth). If the plane surface of undeformed helium is at $z = 0$, then in the deformation case $\bar \xi  < 0$. Increasing $E_{}^\text{(e)}$, increases  $\bar \xi$, which remains flat up to a certain critical value of the electric field $E_{\text c}^\text{(e)}$.
 If $E_{}^\text{(e)} > E_{\text c}^\text{(e)}$, the surface profile acquires a periodic structure. 
 The control parameter for such a phase transition can be not only ${E^\text{(e)}}$ but also the temperature~$T$ and the number of electrons per unit of helium surface area ${n_{\text s}}$ (defined below). Further on, the equation describing a ``critical'' surface relating ${E^\text{(e)}}$, $T$ and ${n_{\text s}}$ at the transition point is obtained.

According to the above described scenario, near the critical point, the surface profile of liquid helium in a phase with lower symmetry has the form~\cite{jmp2012LytvynenkoSK,past2012SlyusarenkoSK,cmp2009LytvynenkoS,jps2015SlyusarenkoL,jpa2017LytvynenkoS}:
\begin{equation}
\label{10a6}
\xi \left( {\boldsymbol{\rho }} \right) = \bar \xi  + \tilde \xi \left( {\boldsymbol{\rho }} \right),
\qquad
\left| {\bar \xi } \right| \gg \left| {\tilde \xi \left( {\boldsymbol{\rho }} \right)} \right|,
\end{equation}
where $\tilde \xi ( {\boldsymbol{\rho}} )$ is a spatially-inhomogeneous  order parameter formed as a result of the phase transition on the background of the flat surface $z = \bar \xi $. In the symmetric phase, $\tilde \xi ( {\boldsymbol{\rho}} ) =0$, while in the asymmetric phase, it describes the spatially periodic structure of the surface. 
In the theory of phase transitions, ``asymmetric phase'' is considered to be the phase (formed due to a phase transition) with the symmetry lower than the initial symmetric phase. If the inequality in equation~\eqref{10a6} takes place in the vicinity of the phase transition point and  $\tilde \xi ( {\boldsymbol{\rho }} )$ vanishes at the point itself, then the second-order phase transition takes place~\cite{book1970Landau5}.

Further on, we consider the case where the surface profile slightly differs from the plane profile and we also assume that $\xi ( {\boldsymbol{\rho}} )$ slowly changes along the coordinates $x$ and $y$, i.e., $| {\partial \xi ({\boldsymbol{\rho}})/\partial x} | \ll 1$ and  $| {\partial \xi ({\boldsymbol{\rho}})/\partial y} | \ll 1$.
If equations~\eqref{12a6},~\eqref{10a6} take place, we can expect that the distribution of electrons and fields will slightly differ from the corresponding distributions in the flat surface case. Thus, the potentials of the total ${\varphi _j}$ and external $\varphi _j^\text{(e)}$ fields ($j = 1,2,3$) can be written in a similar way to equation~\eqref{10a6}, i.e., ${\varphi _j}( {z,{\boldsymbol{\rho}}} ) = {\bar \varphi _j}( z ) + {\tilde \varphi _j}( {z,{\boldsymbol{\rho}}} )$,
where ${\bar \varphi _j}( z )$
are the potentials of the total electric field in ``1''--``3'' regions in the flat surface case. Here, ${\tilde \varphi _j}( {z,{\boldsymbol{\rho}}} )$
are the low perturbations ($| {{{\bar \varphi }_j}( z )} | \gg | {{{\tilde \varphi }_j}( {z,{\boldsymbol{\rho}}} )} |$) of the potentials in all regions caused by the surface inhomogeneity $\tilde \xi ( {\boldsymbol{\rho}} )$.

Let us assume the order parameter $\tilde \xi ({\boldsymbol{\rho}})$ to be spatially periodic.
In this case, applying the averaging over the period $\langle {\ldots} \rangle$ to  equation~\eqref{10a6}, we have
\begin{equation}
\label{19a6}
\bar \xi  \equiv \left\langle {\xi \left( {\boldsymbol{\rho}} \right)} \right\rangle,
\quad
\tilde \xi \left( {\boldsymbol{\rho}} \right) = \xi \left( {\boldsymbol{\rho}} \right) - \left\langle {\xi \left( {\boldsymbol{\rho}} \right)} \right\rangle,
\quad
\tilde \xi \left( {\boldsymbol{\rho}} \right) = \sum\limits_{{\bf{q}} \ne 0} {{\xi _{\bf{q}}}{\re^{\ri{\bf{q}\boldsymbol\rho }}}},
\quad
{\xi _{\bf{q}}} = \frac{1}{{{{\left( {2\piup } \right)}^2}}}\int \rd^2 {\boldsymbol{\rho}}\,\xi \left( {\boldsymbol{\rho}} \right){\re^{ - \ri{\bf{q}\boldsymbol\rho }}}.
\end{equation}

Taking into account equations~\eqref{10a6},~\eqref{19a6}, we also search ${\tilde \varphi _j}( {z,{\boldsymbol{\rho}}} )$ and $\tilde \varphi _j^\text{(e)}( {z,{\boldsymbol{\rho}}} )$ in the form of periodic functions. Therefore, these perturbations also obey the relations similar to~\eqref{19a6}. We do not present them to avoid the aggregation of similar formulae. 
Let us also note that according to equation~\eqref{19a6}, $\langle {\xi ( {\boldsymbol{\rho}} )} \rangle$ is not equal to zero in the most general case. This is due to the skipping helium incompressibility condition during the variation procedure mentioned in the previous section remark.

Since the phase transition is a second-order transition, 
the order parameter $\tilde \xi ( {\boldsymbol{\rho}} )$ can be obtained as a function of the control parameters $T$, ${E^\text{(e)}}$, $n_{\text s}$ near the critical values ${T_{\text c}}$, $E_{\text c}^\text{(e)}$ and ${n_{\text{sc}}}$ by means of the perturbation theory in small parameters $\tilde \xi ( {\boldsymbol{\rho}} )$, ${\tilde \varphi _j}( {z,{\boldsymbol{\rho}}} )$, $\tilde \varphi _j^\text{(e)}( {z,{\boldsymbol{\rho}}} )$, $T - {T_{\text c}}$, ${E^\text{(e)}} - E_{\text c}^\text{(e)}$ and ${n_{\text s}} - {n_{\text{sc}}}$.
Expanding equations~\eqref{5a6}--\eqref{8a6} in these perturbations and averaging them [see equation~\eqref{19a6}],
we obtain the equations describing the system of electrons above a flat helium surface and being a background for the periodic structures research.
To obtain the critical parameters of a phase transition, it is necessary to consider the next orders of the perturbation theory. 
We search ${\tilde \xi _{\bf{q}}}$, ${\tilde \varphi _{j{\bf{q}}}}( z )$ and $\tilde \varphi _{j{\bf{q}}}^\text{(e)}( z )$ in the following form:
\begin{equation}
\begin{gathered}
\label{25a6}
{\tilde \xi _{\bf{q}}}(z) = \sum\limits_{l = 1}^\infty  {\tilde \xi _{\bf{q}}^{\left( l \right)}},
\qquad
\Delta \left( q \right) = \left\{ \begin{array}{ll}
0, & q \ne 0,\\
1, & q = 0,
\end{array} \right.
\\
\tilde \xi _{\bf{q}}^{\left( 1 \right)} = \tilde \xi _{{q_0}}^{\left( 1 \right)}\left[ {\Delta \left( {q - {q_0}} \right) + \Delta \left( {q + {q_0}} \right)} \right],
\qquad
\tilde \xi _{\bf{q}}^{\left( 2 \right)} = \tilde \xi _{2{q_0}}^{\left( 2 \right)}\left[ {\Delta \left( {q - 2{q_0}} \right) + \Delta \left( {q + 2{q_0}} \right)} \right].
\end{gathered}
\end{equation}
The quantities $\tilde \varphi _{j{\bf{q}}}^{( 1 )}( z )$ and $\tilde \varphi _{j{q_0}}^{( 1 )}( z )$, $\tilde \varphi _{j{\bf{q}}}^{( 2 )}( z )$ and $\tilde \varphi _{j2{q_0}}^{( 2 )}( z )$, $\tilde \varphi _{j{\bf{q}}}^{( \text{e} )( 1 )}( z )$ and $\tilde \varphi _{j{q_0}}^{( \text{e} )( 1 )}( z )$, $\tilde \varphi _{j{\bf{q}}}^{( \text{e} )( 2 )}( z )$ and $\tilde \varphi _{j2{q_0}}^{( \text{e} )( 2 )}( z )$ are related in a similar way to equation~\eqref{25a6}.
In equation~\eqref{25a6}, for the purpose of simplicity, the periodic structure is assumed to be one-dimensional with a period $a$ along the $x$-axis, so ${q_0} = {q_{0x}} = {{2\piup } / a}$.
 We do not present here the whole system of self-consistency equations in the basic, first, and other orders of the perturbation theory, because a similar procedure of their obtaining is given in~\cite{jmp2012LytvynenkoSK, jps2015SlyusarenkoL,jpa2017LytvynenkoS}.

It is important to emphasize that the search for a solution for $\tilde \xi ( {\boldsymbol{\rho}} )$ in the 1D case is caused by two circumstances. Firstly, this case simplifies the procedure for obtaining analytical solutions for self-consistency equations~\eqref{5a6},~\eqref{6a6},~\eqref{8a6} and~\eqref{12a6}, because in the case of 2D periodic structures, the calculations become too cumbersome. Secondly, the authors of~\cite{prl1981GianettaI} registered the appearance of similar one-period wave-type structures further evolving to the hexagonal type. The justification for the existence of one-periodic solutions can be a solution to the dynamic stability problem of such structures in the system considered. However, solving this problem is out of the scope of the present paper.

\section{The distribution of electrons and electrostatic fields above flat helium surface}

According to the approach used in~\cite{jmp2012LytvynenkoSK,jps2015SlyusarenkoL,jpa2017LytvynenkoS}, the basic approximation of the first equation in equation~\eqref{8a6} has the form:
\begin{equation}
\label{29a6}
\Delta {{\bar \varphi }_1}\left( z \right)
= 4\piup e
n\left( {{{\bar \varphi }_1}} \right),
\qquad
n\left( {{{\bar \varphi }_1}} \right) = \frac{{\sqrt 2 a_0^{ - 3/2}}}{{{\piup ^2}{e^3}}}\int\limits_0^\infty  {\rd\varepsilon {\varepsilon ^{1/2}}{{\left( {1 + \exp \frac{{\varepsilon  - \psi }}{T}} \right)}^{ - 1}}},
\qquad
\psi = \mu  + e{{\bar \varphi }_1}\left( z \right),
\end{equation}
where 
${a_0} \equiv \frac{{{\hbar ^2}}}{{m{e^2}}}$ is the first Bohr radius and $\psi $ is usually referred to as the electrochemical potential.
Equation~\eqref{29a6} has the form similar to the Thomas-Fermi equation that describes the potential of a many-electron atom in the semiclassical approximation~\cite{book1977Landau3}.
The present paper also uses the semiclassical approximation, and Wigner distribution function ${f_{\bf{p}}}( {\bf{r}} )$ for electrons depends on the coordinate $\bf{r}$ and momentum $\bf{p}$ simultaneously.
The condition for the applicability of this approximation will be obtained at the end of this section. Analogically to the system of electrons in the field of the atom nucleus, in the system of electrons above liquid helium in an external field, the role of the nuclear field is taken by the external clamping field. However, unlike the Thomas-Fermi equation, equation~\eqref{29a6} is obtained from the above formulated variation principle and is of the Cartesian symmetry type rather than of spherical type. Taking into account the noted similarities and differences between the electron system in the field of the atom nucleus and the system electrons above the helium surface in the external clamping field, we can consider equation~\eqref{29a6} as a certain modification of the Thomas-Fermi equation. The solution of both equations requires the application of numerical methods, and the solution of equation~\eqref{29a6} is given below.

The order of equation~\eqref{29a6} can be lowered (see~\cite{jmp2012LytvynenkoSK}):
\begin{equation}
\label{30a6}
\frac{{\partial {{\bar \varphi }_1}}}{{\partial z}} =
-{\left[ {\frac{{{\piup ^{{3 \mathord{\left/
 {\vphantom {3 2}} \right.
 \kern-\nulldelimiterspace} 2}}}{2^{{9 \mathord{\left/
 {\vphantom {9 2}} \right.
 \kern-\nulldelimiterspace} 2}}}{e^2}}}{{3a_0^4}}{{\left( {\frac{{T{a_0}}}{{\piup {e^2}}}} \right)}^{{5 \mathord{\left/
 {\vphantom {5 2}} \right.
 \kern-\nulldelimiterspace} 2}}}\int\limits_0^\infty  {\frac{{\rd x\,{x^{3/2}}}}{{1 + {\re^{x - \chi \left( z \right)}}}}}  + {C_1}} \right]^{1/2}},
\qquad
x = \frac{\varepsilon }{T}\,,
\qquad
\chi \left( z \right) \equiv \frac{{\psi}}{T}.
\end{equation}

Since there are no electrons far from the surface and, hence, in this region the distribution function tends to zero, it is easy to determine the integration constant ${C_1} = E_\infty ^2$, where ${E_\infty } =  - \mathop {\lim }\limits_{z \to  + \infty } {\bar \varphi _1}^\prime ( z )$.
In this case, equation~\eqref{30a6} takes the form:
\begin{equation}
\label{32a6}
{E_1}\left( z \right) = {E_\infty }{\left[ {1 - {2^{{5 \mathord{\left/
 {\vphantom {5 2}} \right.
 \kern-\nulldelimiterspace} 2}}}{\text{Li}_{5/2}}\big( { - {\re^{\,\chi \left( z \right)}}} \big)\frac{{z_0^2}}{{a_0^2}}{{\left( {\frac{{{a_0}T}}{{\piup {e^2}}}} \right)}^{{1 \mathord{\left/
 {\vphantom {1 2}} \right.
 \kern-\nulldelimiterspace} 2}}}} \right]^{1/2}},
\quad
{z_0} = \frac{T}{{e{E_\infty }}}\,,
 \quad
{\text{Li}_s}\left( t \right) = \frac{t}{{\Gamma \left( s \right)}}\int\limits_0^\infty  {\frac{{{x^{s - 1}}\rd x}}{{{\re^x} - t}}}\,,
\end{equation}
where we introduce the polylogarithmic function ${\text{Li}_{s}}( t )$,
that can also be used to rewrite the electron density equation~\eqref{3a6}. Then, integrating it by $z$ from $\bar \xi$ to $ + \infty$, we obtain:
\begin{equation}
\label{36a6}
4 \piup e{n_{\text s}} = {E_0} - {E_\infty }\,,
\qquad
{n_{\text s}} = \int\limits_\xi ^\infty  {\rd z\,n( z )},
\qquad
n( z ) =  - {\left( {\frac{{T{a_0}}}{{\piup {e^2}}}} \right)^{{3 \mathord{\left/
 {\vphantom {3 2}} \right.
 \kern-\nulldelimiterspace} 2}}}\frac{{{\text{Li}_{{3 \mathord{\left/
 {\vphantom {3 2}} \right.
 \kern-\nulldelimiterspace} 2}}}\big( { - {\re^{\,\chi \left( z \right)}}} \big)}}{{\sqrt 2 a_0^3}}\,,
\end{equation}
where ${E_0} = { { - {{\bar \varphi }_1}^\prime ( z )} |_{z = \bar \xi }}$ 
and ${n_{\text s}}$ is the total number of electrons in the system above the surface area unit. Equation~\eqref{36a6} together with equation~\eqref{32a6} at $z = \bar \xi$
give an implicit equation for obtaining the non-dimensionalized electrochemical potential on the liquid helium surface ${\chi _\xi } = \chi ( {z = \bar \xi } )$ as a function of $T$, ${E^\text{(e)}}$ and $n_{\text s}$:
\begin{equation}
\label{38a6}
\frac{{{E_\infty }}}{{4\piup e}}\left\{ {{{\left[ {1 - {2^{{5 \mathord{\left/
 {\vphantom {5 2}} \right.
 \kern-\nulldelimiterspace} 2}}}\frac{{z_0^2}}{{a_0^2}}{{\left( {\frac{{T{a_0}}}{{\piup {e^2}}}} \right)}^{{1 \mathord{\left/
 {\vphantom {1 2}} \right.
 \kern-\nulldelimiterspace} 2}}}\text{Li}_{5/2}\left( { - {\re^{{\,\chi _\xi }}}} \right)} \right]}^{1/2}} - 1} \right\} = {n_{\text s}}.
\end{equation}
On the other hand, equation~\eqref{38a6} is also a condition for normalizing the electrochemical potential at fixed values of  $T$, ${E^\text{(e)}}$ and $n_{\text s}$, if the condition $\text{Li}_{5/2}( { - {\re^{{\,\chi _\infty }}}} ) = 0$ is satisfied. The last condition is satisfied due to the absence of electrons at infinity. This fact was also used by us to determine the integration constant in equation~\eqref{30a6}.

To obtain the relation between $E_1^\text{(e)}$ and  $E_1^{}( z ) = E_1^{\text{(i)}}( z ) + E_1^\text{(e)}$ we must obtain the electric field $E_1^{\text{(i)}}( z )$ of electrons with density equation~\eqref{3a6}. The
 $z$-component of $E_1^{\text{(i)}}( z )$ at point $( {x,y,z} )$ induced by an elementary volume $\rd x'\rd y'\rd z'$ of electrons at point $( {x',y',z'} )$ is $\rd E_{1z}^{\text{(i)}}( z ) =  - ( {z - z'} )en( {z'} )\rd x'\rd y'\rd z' \times \allowbreak  {[ {{{x'}^2} + {{y'}^2} + {{( {z - z'} )}^2}} ]
 ^{{{ - 3} / 2}}}$.
Since the system is infinite along $( {x,y} )$ coordinates, the integrals of the projections $E_{1x}^{\text{(i)}},E_{1y}^{\text{(i)}}$ over $V_1$ volume vanish. 
Thus, $E_1^{\text{(i)}}( z )$ can be calculated by integrating 
 $\rd E_{1z}^{\text{(i)}}( z )$ over the volume above the helium surface $E_{1z}^{\text{(i)}}( z ) =
 \int\nolimits_{{V_1}} {\rd E_{1z}^{\text{(i)}}( z )}
 =- 2\piup e[ {\int\nolimits_{\bar \xi }^z {\rd z'n( {z'} ) - } \int\nolimits_z^\infty  {\rd z'n( {z'} )} } ]$.
The calculation of the integral in $E_1^{\text{(i)}}( z )$ requires the application of numerical methods, but considering its limit cases at $z \to \bar \xi$ and $z \to  + \infty$, can give a simple physical interpretation of $E_1^{\text{(i)}}( z )$.
In the first case,
we get $E_{1z}^{\text{(i)}}( {z = \bar \xi } ) = 2\piup e{n_{\text s}}$. In the second case,
 $E_{1z}^{\text{(i)}}(z \to  + \infty )= - 2\piup e{n_{\text s}}$.
Thus, the field induced by electrons on both sides from their location (i.e., in the first case, all electrons are located above the observation point, in the second --- below) is equivalent to the field of a charged plate with the surface charge density $e{n_{\text s}}$. 
 ${E^\text{(e)}} $ can be obtained, e.g., as the difference
 $ \mathop {\lim }\nolimits_{z \to  + \infty } [ {{E_1}( z ) - E_1^{\text{(i)}}( z )} ] = E_1^\text{(e)} = {E_\infty } + 2\piup e{n_{\text s}}$.
In quasi-neutral case at $z \to  + \infty$,
the external pressing field $E^\text{(e)}$ is compensated by the field of electrons
$E_1^{\text{(i)}}( z )$, so,
 ${E_\infty } = 0$, 
and we obtain the relation coinciding with the result of paper~\cite{jpa2017LytvynenkoS}:
\begin{equation}
\label{44a6}
E_1^\text{(e)} = 2\piup e{n_{\text s}}.
\end{equation}
As ${E_1}( {z > \bar \xi } ) > 0$
[see equations~\eqref{30a6},~\eqref{32a6}], we have the condition $E_1^\text{(e)} \geqslant 2\piup e{n_{\text s}}$ (otherwise, \linebreak $\mathop {\lim }\nolimits_{z \to  + \infty } E_1^{}( z ) $ 
may be less than zero).
Let us remind that for the quasi-neutral case we consider the situation where the external clamping field $E_1^\text{(e)}$ is compensated by the field of electrons $E_1^{\text{(i)}}$ at a substantial distance from the helium surface.

According to equation~\eqref{32a6}, the expression for $\chi ( z )$ can be obtained in quadrature and the solution for the main approximation of equation~\eqref{8a6} with the boundary conditions of equation~\eqref{12a6} of the perturbation theory has the form:  
\begin{equation}
\begin{gathered}
\label{46a6}
\int\limits_{{{\chi _\xi }}}^\chi  {\rd\chi '{{\left[ {1 - {2^{{5 \mathord{\left/
 {\vphantom {5 2}} \right.
 \kern-\nulldelimiterspace} 2}}}\text{Li}_{5/2}\big( { - {\re^{\,\chi '}}} \big)\frac{{z_0^2}}{{a_0^2}}{{\left( {\frac{{{a_0}T}}{{\piup {e^2}}}} \right)}^{{1 \mathord{\left/
 {\vphantom {1 2}} \right.
 \kern-\nulldelimiterspace} 2}}}} \right]}^{ - 1/2}}}  = \frac{{\bar \xi  - z}}{{{z_0}}}\,,
 \qquad
 {\bar \varphi _2}\left( z \right) =  - \frac{{{E_0}}}{\varepsilon }\left( {z - \bar \xi } \right) + {\varphi _0}\,,
 \\
 {\bar \varphi _3}\left( z \right) =  - \frac{{{E_0}}}{{{\varepsilon _d}}}\left( {z + d} \right) + \frac{{{E_0}}}{\varepsilon }\left( {d + \bar \xi } \right) + {\varphi _0}\,,
 \qquad
 \bar \varphi _1^\text{(e)}(z) =  - E_1^\text{(e)}\left( {z - \bar \xi } \right) + \varphi _0^\text{(e)},
 \\
 \bar \varphi _2^\text{(e)}(z) =  - \frac{{E_1^\text{(e)}}}{\varepsilon }\left( {z - \bar \xi } \right) + \varphi _0^\text{(e)},
 \qquad
 \bar \varphi _3^\text{(e)}(z) =  - \frac{{E_1^\text{(e)}}}{{{\varepsilon _d}}}\left( {z + d} \right) + \frac{{E_1^\text{(e)}}}{\varepsilon }\left( {d + \bar \xi } \right) + \varphi _0^\text{(e)},
\end{gathered}
\end{equation}
where 
${\bar \varphi _{1 }}( {z = \bar \xi } ) 
\equiv {\varphi _0}$ and $\bar \varphi _{1 }^\text{(e)}( {z = \bar \xi } ) 
\equiv \varphi _0^\text{(e)}$. 

Based on equations~\eqref{5a6},~\eqref{38a6},~\eqref{46a6}, the surface level subsidence $\bar \xi$ in terms of the problem parameters has the form:
\begin{equation}
\label{47a6}
\bar \xi  =  - \frac{{{{\left( {4\piup e{n_{\text s}}} \right)}^2}}}{{8\piup \alpha {\kappa ^2}}}\left[ {\frac{{E_{}^\text{(e)}}}{{2\piup e{n_{\text s}}}}\left( {1 + \frac{1}{{2\varepsilon }}} \right) + \frac{1}{{4\varepsilon }}} \right].
\end{equation}

As seen from equation~\eqref{47a6}, an increase of the values of $E_{}^\text{(e)}$ and ${n_{\text s}}$  may cause the breaking of the natural condition:
\begin{equation}
\label{48a6}
\left| {\bar \xi } \right| < d.
\end{equation}
Obviously, the value of lowering of the helium surface level $| {\bar \xi } |$ leads to the film thickness value $d - | {\bar \xi } |$. Thus, when calculating the equation~\eqref{47a6}, the value of $\kappa ( d )$ should be replaced by $\kappa ( {d - | {\bar \xi } |} )$.
This  takes into account the effect of lowering of the surface level $\bar \xi ( {d,{n_{\text s}},E_{}^\text{(e)}} )$ correctly, but its obtaining requires using numerical methods. Numerical estimates of equation~\eqref{47a6} show that for $E_{}^\text{(e)} = 5000$~V/cm, ${n_{\text s}} = 5 \cdot {10^8}$~cm$^{ - 2}$, $T = 5$~K and $d = 0.1$~cm condition~\eqref{48a6} takes place even in the case of strong inequality $| {\bar \xi } | \ll d$. Taking into account the experimental data of \cite{pla1979LeidererW,ss1982LeidererES,pla1980EbnerL}, where the values of $E_{}^\text{(e)}$, $n_{\text s}$ and $T$ are lower than the above mentioned, and the value of $d$ is higher, the results obtained in the present paper can be compared with the experimental data using equation~\eqref{47a6}.

\begin{figure}[!t]
\centering
\subfigure[Quasi-neutral system, $E_{}^\text{(e)} = 2\piup e{n_{\text s}}$]{
\includegraphics[width=.48\textwidth]{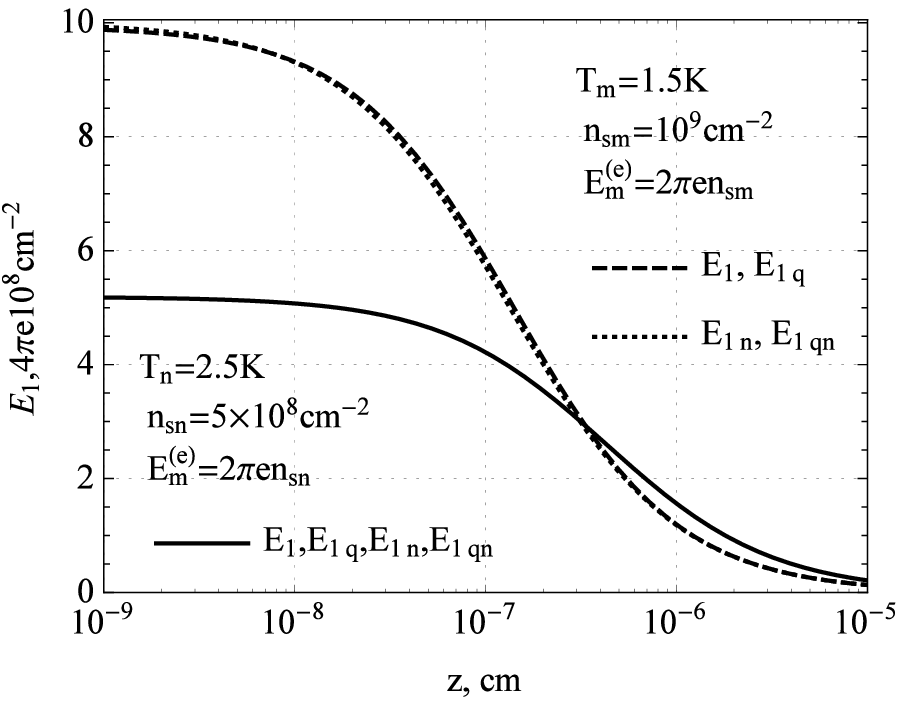}
}
\subfigure[Charged system, $E_{}^\text{(e)} > 2\piup e{n_{\text s}}$]{
\includegraphics[width=.48\textwidth]{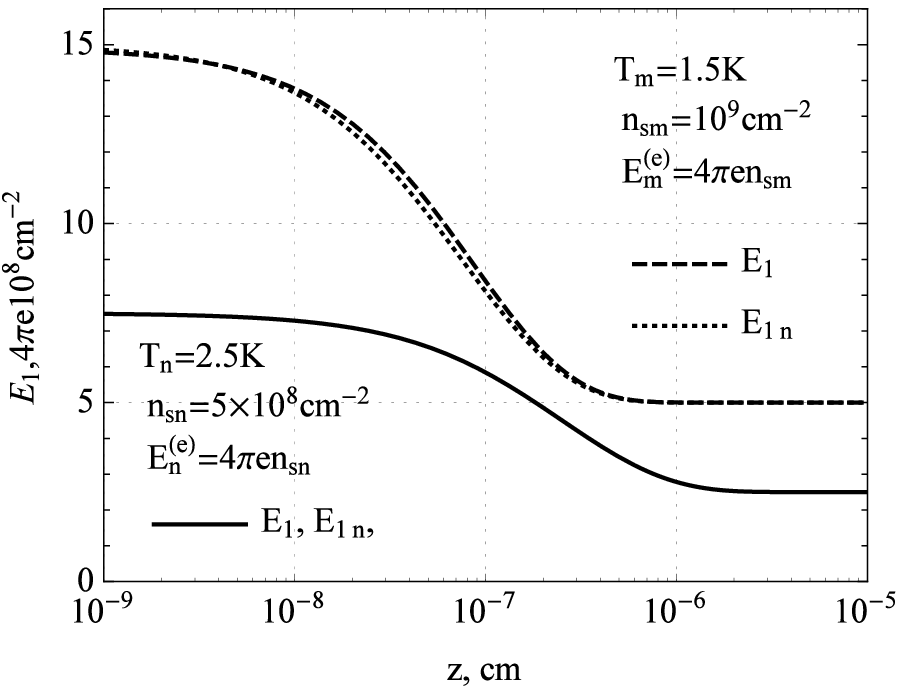}
}
\caption{$E_1( z )$  for different $T$, $n_{\text s}$ and ${E^\text{(e)}}$  values:
 ${\Delta _\text{n}}( {T_\text{n},E_\text{n}^\text{(e)}=2\piup e{n_{\text{sn}}},{n_{\text{sn}}}} ) \approx 0.05$,
 ${\Delta _\text{n}}( {T_\text{n},E_\text{n}^\text{(e)}=} $ \allowbreak
 ${4\piup e{n_{\text{sn}}},{n_{\text{sn}}}} ) \approx 0.099$,
 ${\Delta _\text{n}}( {T_{\text m}^{},E_{\text m}^\text{(e)}=2\piup e{n_\text{sm}},{n_\text{sm}}} ) \approx 0.72$,
  ${\Delta _\text{n}}( {T_{\text m}^{},E_{\text m}^\text{(e)}=4\piup e{n_\text{sm}},{n_\text{sm}}} ) \approx 1.44$.
}
\label{fig:2a6}
\end{figure}

In paper~\cite{jpa2017LytvynenkoS}, the problem of describing the distribution of charges (electrons) above the liquid dielectric surface in the quasi-neutral case ($E_{}^\text{(e)} = 2\piup e{n_{\text s}}$) was considered. Paper~\cite{jmp2012LytvynenkoSK} considered a system of a non-degenerate gas of charges (electrons) above the liquid dielectric surface in the ``charged'' case ($E_{}^\text{(e)} > 2\piup e{n_{\text s}}$). The present paper can be considered as a generalization of these two articles~\cite{jmp2012LytvynenkoSK,jpa2017LytvynenkoS}. Therefore, the results of these papers can be obtained as the limit cases of equations~\eqref{32a6},~\eqref{36a6},~\eqref{46a6} and~\eqref{47a6}. In particular, by setting $E_{}^\text{(e)} = 2\piup e{n_{\text s}}$, we can obtain the results of the quasi-neutral problem~\cite{jpa2017LytvynenkoS}, e.g., the electric field ${E_{1q}}( z ) = \frac{{T{2^{{5 /4}}}}}{{e{a_0}}}{( {\frac{{T{a_0}}}{{\piup {e^2}}}} )^{{1 /4}}}{[ { - \text{Li}_{{5 / 2}}( { - {\re^{\,\chi ( z )}}} )} ]^{{1 /2}}}$.
In order to carry out the limit case of a non-degenerate gas of electrons, let us consider the range of $T$, ${E^\text{(e)}}$ and $n_{\text s}$ values, where the particle distribution function equation~\eqref{6a6} is close to Boltzmann's distribution, i.e., ${\re^{\frac{{\varepsilon  - \psi }}{T}}} \gg 1$.
This inequality allows one to obtain the main approximation of an arbitrary order polylogarithmic function $\text{Li}_s( { - {\re^{\,\chi ( z )}}} ) \approx  - {\re^{\,\chi} }$ [see equation~\eqref{32a6}] and integrate the first equation in equation~\eqref{46a6}, so ${\chi _\text{n}}( z )$ can be obtained (``n'' subscript marks the non-degeneracy case).
Using equations~\eqref{32a6},~\eqref{36a6},~\eqref{47a6}, 
we can calculate the density ${n_\text{n}}( z )$  and electric field ${E_{1\text{n}}}( z )$ distributions as well.
Doing this and taking into account equations~\eqref{6a6},~\eqref{38a6} allows one to obtain the non-degeneracy region of the gas of electrons in $\{ {T,E,{n_{\text s}}} \}$ space:
\begin{equation}
\label{63a6}
{\Delta _\text{n}} \ll 1,
 \qquad
{\Delta _\text{n}} = {2^{{1 \mathord{\left/
 {\vphantom {1 2}} \right.
 \kern-\nulldelimiterspace} 2}}}{\piup ^{{3 \mathord{\left/
 {\vphantom {3 2}} \right.
 \kern-\nulldelimiterspace} 2}}}{n_{\text s}}{e^4}{E^\text{(e)}}a_0^{{3 \mathord{\left/
 {\vphantom {3 2}} \right.
 \kern-\nulldelimiterspace} 2}}{T^{{{ - 5} \mathord{\left/
 {\vphantom {{ - 5} 2}} \right.
 \kern-\nulldelimiterspace} 2}}}.
\end{equation}
Using the approximation in equation~\eqref{63a6}, we obtain the expression for the electric field ${E_{1\text{n}}}( z ) = {E_\infty }\frac{{1 + X( z )}}{{1 - X( z )}}$, where $ X( z ) \equiv \frac{{{E_0} - {E_\infty }}}{{{E_0} + {E_\infty }}}\exp ( {\frac{{\bar \xi  - z}}{{{z_0}}}} )$, which corresponds to the results of~\cite{jmp2012LytvynenkoSK}. Considering the approximation in equation~\eqref{63a6} in quasi-neutral case (${E^\text{(e)}} \to 2\piup e{n_{\text s}}$), we obtain the results of~\cite{jps2015SlyusarenkoL}, where a non-degenerate gas of charges (electrons) in the quasi-neutral case was considered, i.e., for the electric field ${E_{1\text{nq}}}( z ) = \frac{{4\piup e{n_{\text s}}}}{{1 + {{( {z - \bar \xi } )} / {( {2{z_{0\text{n}}}} )}}}}$ (``nq'' index marks such a case). Figure~\ref{fig:2a6} shows  the relation between ${E_{1}}( z )$, ${E_{1\text{q}}}( z )$, ${E_{1\text{n}}}( z )$ and ${E_{1\text{nq}}}( z )$.

Concluding this section, let us obtain the condition of the semiclassical approximation applicability considered in the present paper. 
Let us calculate the mean de~Broglie heat wave-length of electrons in the system as
$\langle \lambda  \rangle$ having the order of $\langle \lambda  \rangle  \sim \hbar  \cdot {\langle {{p^2}} \rangle ^{{{ - 1} / 2}}}$, where  $\langle {{p^2}} \rangle$ is the averaged squared momentum (see below).
If the distance between electrons is much smaller than $\langle \lambda  \rangle$, the quasi-classical approximation fails.
Assuming that the average distance between electrons near the helium surface is of the order of ${( {{{ n |}_{z = \bar \xi }}} )^{ - {1 / 3}}}$, we obtain the condition for the quasi-classical approximation applicability:
\begin{equation}
\label{70a6}
{\Delta _\lambda }=\left\langle \lambda  \right\rangle  \cdot {\left( {{{\left. n \right|}_{z = \bar \xi }}} \right)^{ - {1 \mathord{\left/
 {\vphantom {1 3}} \right.
 \kern-\nulldelimiterspace} 3}}},
\quad
\left\langle \lambda  \right\rangle  \sim \hbar  \cdot {\left\langle {{p^2}} \right\rangle ^{{{ - 1} \mathord{\left/
 {\vphantom {{ - 1} 2}} \right.
 \kern-\nulldelimiterspace} 2}}},
 \quad
\left\langle {{p^2}} \right\rangle  = {{\int {{\rd^3}{\bf r}\,{\rd^3}{\bf p}} {f_{\bf{p}}}\left( {\bf{r}} \right){p^2}} \mathord{\left/{\vphantom {{\int {{\rd^3}r{\rd^3}p} {f_{\bf{p}}}\left( {\bf{r}} \right){p^2}} {\int {{\rd^3}\bf r\,{\rd^3}\bf p} {f_{\bf{p}}}\left( {\bf{r}} \right)}}} \right.\kern-\nulldelimiterspace} {\int {{\rd^3}{\bf r}\,{\rd^3}{\bf p}} {f_{\bf{p}}}\left( {\bf{r}} \right)}}.
\end{equation}
Further on, (see table~\ref{tbl-smp1}), it is shown that
the obtained results for the critical parameters of the phase transition are used in the region of $T$, $E^\text{(e)}$ and ${n_{\text s}}$ values, where equation~\eqref{70a6} takes place.

\section{Critical parameters of the phase transition in the system to a spatially periodic state}

The initial point for the research of the critical parameters of the phase transition is to obtain the relation between $\tilde \varphi _j^{( 1 )}$ and the order parameter $\tilde \xi _{}^{( 1 )}$.
Following the similar procedure from ~\cite{jmp2012LytvynenkoSK,jps2015SlyusarenkoL,jpa2017LytvynenkoS}, we can obtain the first harmonics of the Fourier transform of the first approximation of equation~\eqref{8a6}:
\begin{equation}
\label{71a6}
\frac{{{\partial ^2}\tilde \varphi _j^{\left( 1 \right)}}}{{\partial {z^2}}} - q_0^2\tilde \varphi _j^{\left( 1 \right)} = 4\piup {e^2}\frac{{\partial n}}{{\partial \mu }}\tilde \varphi _j^{\left( 1 \right)}{\delta _{j1}}\,,
\quad
\frac{{{\partial ^2}\tilde \varphi _j^{\left( \text{e} \right)\left( 1 \right)}}}{{\partial {z^2}}} = q_0^2\tilde \varphi _j^{\left( \text{e} \right)\left( 1 \right)}, \quad
j = 1,2,3,
\quad
{\delta _{ij}} = \left\{ \begin{array}{l}
1,\;i = j\\
0,\;i \ne j
\end{array} \right.
.
\end{equation}
Further on, only some of the solutions $\tilde \varphi _j^{( 1 )}$ and $\tilde \varphi _j^{( \text{e} )( 1 )}$ for equation~\eqref{71a6} are needed (see \cite{jmp2012LytvynenkoSK,jps2015SlyusarenkoL,jpa2017LytvynenkoS} for the details). Considering the approximation ${\Delta _1} \gg 1$, where ${\Delta _1} = 4\piup \frac{{{e^2}}}{{q_0^2}}| {\frac{{\partial n}}{{\partial \mu }}} |$, the solutions satisfying the boundary conditions equation~\eqref{12a6}, have the form:
\begin{equation}
\begin{gathered}
\label{83a6}
{\tilde \varphi ^{\left( 1 \right)}}_1 = \tilde \xi _{}^{\left( 1 \right)}{E_1}\left( z \right)G\left( {{q_0}} \right),
\quad
\tilde \varphi _2^{\left( 1 \right)}\left( z \right) = \tilde \xi _{}^{\left( 1 \right)}{E_0}\eta \left( z \right)F\left( {{q_0}} \right),
\quad
\tilde \varphi _2^{\left( \text{e} \right)\left( 1 \right)}\left( z \right) =  - \tilde \xi _{}^{\left( 1 \right)}\eta \left( z \right)E_1^\text{(e)}{F^\text{(e)}}\left( {{q_0}} \right),
\\
\eta \left( z \right) = {\re^{{q_0}\left( {z - \bar \xi } \right)}} - C{\re^{{q_0}\left( {\bar \xi  - z} \right)}},
\quad
C = \frac{{{\varepsilon _d} - \varepsilon }}{{{\varepsilon _d} + \varepsilon }}{\re^{ - 2{q_0}\left( {d + \bar \xi } \right)}},
\quad
{F^\text{(e)}}\left( {{q_0}} \right) = \frac{{1 - {\varepsilon ^{ - 1}}}}{{\varepsilon \left( {1 + C} \right) + 1 - C}}\,,
\\
G\left( {{q_0}} \right) = \frac{{{{\bar \varphi ''}_1}\left( {1 - C} \right) + \left( {\varepsilon  - 1} \right){E_0}{q_0}\left( {1 + C} \right)}}{{{{\bar \varphi ''}_1}\left( {1 - C} \right) + {E_0}\varepsilon {q_0}\left( {1 + C} \right)}}\,,
\quad
F\left( {{q_0}} \right) = \frac{{{{{{\bar \varphi ''}_1}} \mathord{\left/
 {\vphantom {{{{\bar \varphi ''}_1}} \varepsilon }} \right.
 \kern-\nulldelimiterspace} \varepsilon }}}{{{{\bar \varphi ''}_1}\left( {1 - C} \right) + {E_0}\varepsilon {q_0}\left( {1 + C} \right)}}.
\end{gathered}
\end{equation}
Substituting equation~\eqref{83a6} into the first harmonics of the Fourier transform of the first approximation of equation~\eqref{5a6} and assuming $\tilde \xi _{}^{( 1 )} \ne 0$, we have (see \cite{jmp2012LytvynenkoSK,jps2015SlyusarenkoL,jpa2017LytvynenkoS} for the details):
\begin{equation}
\begin{gathered}
\label{89a6}
{q_0}\left( {C + 1} \right)\left[ {\left( {1 + \varepsilon } \right)E_0^2F\left( {{q_0}} \right) + E{{_1^\text{(e)}}^2}{F^\text{(e)}}\left( {{q_0}} \right)} \right] - 4\piup \alpha {\kappa ^2}\left[ {1 + q_0^2{\kappa ^{ - 2}}\left( {1 + {\kappa ^2}{{{{\bar \xi }^{\,2}}} / 2}} \right)} \right] = 0
.
\end{gathered}
\end{equation}
This equation describes a critical surface in 
space $\{ {T,{E^\text{(e)}},{n_{\text s}}} \}$ as well as the modulus of reciprocal lattice vector ${q_0}$  of the periodic structure that appeared.
This means that in the most general case, the phase transition can be considered  regarding three critical parameters $E_{\text c}^\text{(e)}( {T,{E^\text{(e)}},{n_{\text s}}} )$, ${T_{\text c}}( {T,{E^\text{(e)}},{n_{\text s}}} )$ and ${n_{\text{sc}}}( {T,{E^\text{(e)}},{n_{\text s}}} )$. However, the current paper is focused on  considering the phase transition occurring at $E^\text{(e)}>E_{\text c}^\text{(e)}$. This option is chosen for the purpose of comparing the obtained results with the experimental data~\cite{pla1979LeidererW,ss1982LeidererES},where the dimple crystals were detected, when the external electric field $E^\text{(e)}$ exceeded a certain critical value $E_{\text c}^\text{(e)}$.

\begin{figure}[!b]
\centering
\subfigure[Phase curves for fixed $q_0$ and $n_{\text s}$]{
\includegraphics[width=.46\textwidth]{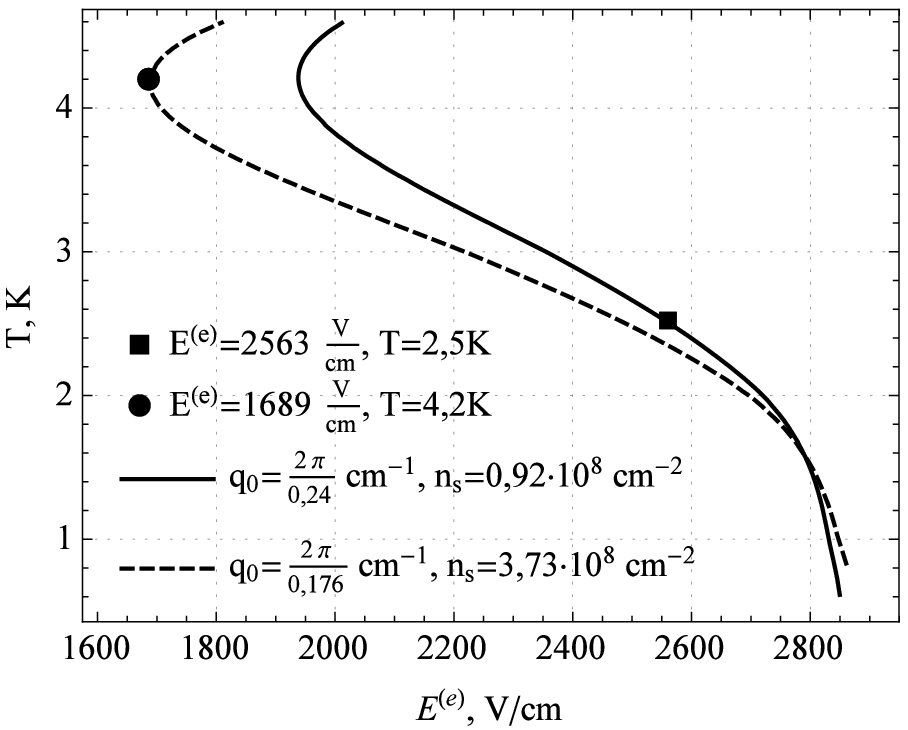}
}
\subfigure[ ${\tilde \xi ^{\left( 1 \right)}}$ near the critical point $E_{\text c}^\text{(e)}$ for fixed $T$, $n_{\text s}$]{
\includegraphics[width=.51\textwidth]{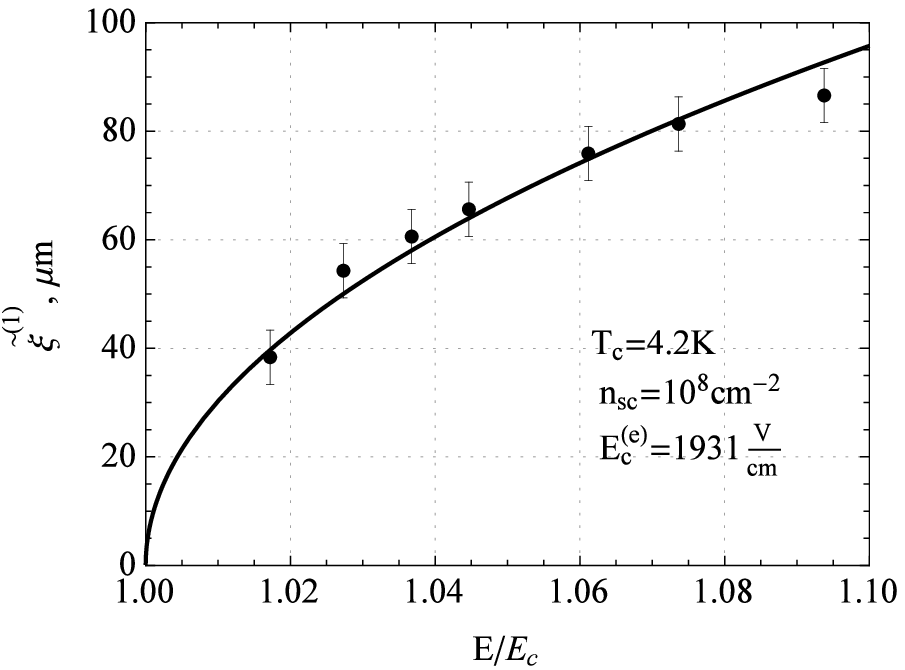}
}
\caption{(a) --- phase curves equation~\eqref{89a6}, (b) --- order parameter  equation~\eqref{90a6}. Experimental data: $ \bullet$~---~\cite{pla1979LeidererW}, $\blacksquare$ --- \cite{ss1982LeidererES}.}
\label{fig:7a6}
\end{figure}

Equation~\eqref{89a6} is in good agreement with the experimental data~\cite{pla1979LeidererW,ss1982LeidererES} (see figure~\ref{fig:7a6} and table~\ref{tbl-smp1}) as ${\Delta _{{E_{\text c}}}} = {{[ {E_{\text c}^{( \text{e} )\exp } - E_{\text c}^\text{(e)}} ]} / {E_{\text c}^{( \text{e} )\exp }}}<0.06$, where ${E_{\text c}^{( \text{e} )\exp }}$ is the experimental value of the clamping field and ${E_{\text c}^\text{(e)}}$ is calculated from equation~\eqref{89a6} using the set of experimental values ${T^{\exp }}$, $n_{\text s}^{\exp }$, ${q_0} = {{2\piup }/ {{a^{\exp }}}}$ and ${d^{\exp }}$. If $n_{\text s}^{\exp }$ is not given directly (e.g., \cite{ss1982LeidererES}), it can be estimated using the number of particles per dimple  $N_1^{\exp }$ and the distance ${a^{\exp }}$ between the dimples. Assuming the elementary cell of hexagonal lattice to be a rhombus with ${a^{\exp }}$ side and ${\piup  / 3}$ internal angle, we have $n_{\text s}^{\exp } \approx {{N_1^{\exp }} / {( {{a^{\exp }}^2\sin {\piup  / 3}} )}}$.
For both \cite{pla1979LeidererW,ss1982LeidererES} parameter values ${\Delta _1} \sim {10^{11}}$, so the approximation used to obtain equation~\eqref{83a6} is valid. The quasi-classical approximation is also valid, because the inequality ${\Delta _\lambda } \ll 1$ takes place [see equation~\eqref{70a6} and table~\ref{tbl-smp1}].

\begin{table}[!t]
\caption{Comparison of experimental critical value $E_{\text c}^{\left( \text{e} \right)\exp }$ with $E_{\text c}^{\left( \text{e} \right) }$ calculated from equation~\eqref{89a6} using experimental data~\cite{pla1979LeidererW,ss1982LeidererES}. Experimental quantities are marked by ``exp'' subscript.  
 }
\label{tbl-smp1}
\vspace{2ex}
\begin{center}
\renewcommand{\arraystretch}{0}
\small
\begin{tabular}{|c|c|c|c|c|c|c|c|c|c|}
\hline\hline
Ref.
&$E_{\text c}^{\left( \text{e} \right) \exp  },~\frac{\text{V}}{{\text{cm}}}
$
&${T^{\exp }}$,~K
&$n_{\text s}^{\exp }$,~cm$^{ - 2}$
&$N_1^{\exp }$
&${a^{\exp }},$~cm
&${d^{\exp }},$~cm
&$E_{\text c}^\text{(e)},~\frac{\text{V}}{{\text{cm}}}
 $
&${\Delta _{{E_{\text c}}}}$
&${\Delta _\lambda }$\strut\\
\rule{0pt}{2pt}&&&&&&&&&\\
\hline\hline
      \cite{pla1979LeidererW}&$1790 \pm 40$&$4.2$&$3.73 \cdot {10^8}$&${10^7}$&$0.176$&$>0.2$&$1689$&  $0.056$&$0.094$\strut\\
\hline
      \cite{ss1982LeidererES}&$2600$&$2.5$&$0.92 \cdot {10^8}$&$5 \cdot {10^6}$&$0.24$&$>0.1$&$2563$&$0.014$&$0.107$\strut\\
\hline\hline
\end{tabular}
\renewcommand{\arraystretch}{1}
\end{center}
\end{table}

Expanding the self-consistency equation system in the small perturbations near the phase transition point up to the next non-vanishing order and calculating its Fourier transform at $q ={q_0}$, we obtain the expression for the order parameter ${\tilde \xi ^{( 1 )}}$:
\begin{equation}
\begin{gathered}
\label{90a6}
{\tilde \xi ^{\left( 1 \right)}} = \gamma \sqrt {\frac{{\partial n}}{{\partial {E^\text{(e)}}}}\left[ {{E^\text{(e)}} - E_{\text c}^\text{(e)}} \right] + \frac{{\partial n}}{{\partial T}}\left( {T - {T_{\text c}}} \right) + \frac{{\partial n}}{{\partial {n_{\text s}}}}\left( {{n_{\text s}} - {n_{\text{sc}}}} \right)},
\end{gathered}
\end{equation}
where $\gamma$ has a rather cumbersome dependence on $T_{\text c}$, $n_{\text{sc}}$, $E_{\text c}^\text{(e)}$ and $q_0$, so we do not present its expression.
Using the experimental data of~\cite{ss1982LeidererES} and fixing ${T_{}} = T_{\text c}$ and ${n_{\text s}=n_{\text{sc}}}$, we can see that ${\tilde \xi ^{( 1 )}} \sim \sqrt {{{{E^\text{(e)}}} /{E_{\text c}^\text{(e)} - 1}}}$ near ${E_{\text c}^\text{(e)}}$ [see figure~\ref{fig:7a6} (b)].
  The first harmonic of the Fourier transform of the density perturbation  ${n^{( 1 )}} = \frac{{\partial n}}{{\partial {\varphi _1}}}{\tilde \varphi _1}^{( 1 )} + \frac{{\partial n}}{{\partial \xi }}{\tilde \xi ^{( 1 )}}$ is of the form:
\begin{equation}
\begin{gathered}
\label{100a6}
{n^{\left( 1 \right)}} =  - T\frac{{\partial n}}{{\partial \mu }}\left[ {1 - G\left( {{q_0}} \right)} \right]\frac{{{{\tilde \xi }^{\left( 1 \right)}}}}{{{z_0}}}.
\end{gathered}
\end{equation}
Numerical estimates show that $G( {{q_0}} ) < 1$, see equation~\eqref{83a6}, so the density maximum corresponds to the dimple on the helium surface and vice versa.

\section{Conclusion}

Thus, in this paper we propose a quantum-statistical theory of equilibrium spatially inhomogeneous states in the system of electrons above the liquid helium surface in an external constant electric clamping field. 
In terms of the built theory, the self-consistency equations~\eqref{5a6},~\eqref{6a6},~\eqref{8a6}  describing the system are obtained. 
 The obtained agreement of the results of the present paper and papers \cite{jmp2012LytvynenkoSK,jps2015SlyusarenkoL,jpa2017LytvynenkoS} is demonstrated in figure~\ref{fig:2a6},
 figure~\ref{fig:7a6} and table~\ref{tbl-smp1}. Going beyond the Boltzmann statistics, the obtained self-consistency equations are also used to describe spatially periodic structures formed as a result of the phase transition near the critical point. 
 The approach also allows one to obtain the phase surface equation equation~\eqref{89a6} that relates ${E^\text{(e)}},T,{n_{\text s}}$ and ${q_0}$ parameters at the phase transition point. 
 The comparison between the obtained results and the available experimental data~\cite{pla1979LeidererW,ss1982LeidererES} gives a good agreement within the experimental measurement error (see figure~\ref{fig:7a6}). It is shown that the nature of the periodic structures~\eqref{100a6} is similar to the experimentally observed one (e.g.,~\cite{pla1979LeidererW,ss1982LeidererES,pla1980EbnerL}).

The proposed approach can be improved in two aspects at least. Firstly, it should be generalized to the case of 
two reciprocal (or direct) lattice vectors characterizing the periodic structures. As noted above, in the experiments~\cite{prl1981GianettaI}, wavy structures were observed as an intermediate state in the transition from spatially-homogeneous to ``2D'' spatially-periodic states. Secondly, 
quantum effects can be taken into account, e.g., the exchange interaction. Currently the authors are working on the both issues.

\ukrainianpart

\title{Про просторово-періодичне упорядкування у системі електронів над поверхнею рідкого гелію у зовнішньому електричному полі}
\author{Д.М. Литвиненко\refaddr{label1,label2}, Ю.В. Слюсаренко\refaddr{label1,label2}, А.І. Кірдін\refaddr{label1}}
\addresses{
\addr{label1}
ННЦ ХФТІ, вул. Академічна, 1, 61108 Харків, Україна
\addr{label2}
ХНУ ім. В.Н. Каразіна, пл. Свободи, 4, 61108 Харків, Україна
}

\makeukrtitle

\begin{abstract}
\tolerance=3000%
Побудовано теорію рівноважних станів електронів над поверхнею рідкого гелію у зовнішньому притискаючому полі на основі перших принципів квантової статистики для систем багатьох тотожних Фермі частинок. В основу підходу покладено варіаційний принцип, модифікований для розглянутих систем, і модель Томаса-Фермі. В термінах розвиненої теорії отримані рівняння самоузгодження, що пов'язують параметри опису такої системи --- потенціал статичного електричного поля, функцію розподілу зарядів і профіль поверхні рідкого діелектрика. Ці рівняння використано для вивчення фазового перетворення системи до просторово-періодичних станів. Як приклад можливостей запропонованого методу, аналізуються характеристики фазового перетворення системи до просторово-періодичних станів жолобкового типу.
\keywords електрони, газо-рідинні границі, варіаційний підхід, теорія збурень, фазові переходи

\end{abstract}


\begin{thebibliography}{35}
\bibitem{pla1971CrandallW} Crandall~R.S., Williams~R., Phys. Lett. A, 1971, \textbf{34}, 404--405, \doi{10.1016/0375-9601(71)90938-8}.
	
\bibitem{jetpl1974Shikin} Shikin~V., JETP Lett., 1974, \textbf{19}, No. 10, 335--336. 

\bibitem{jetp1975MonarkhaS} Monarkha~Yu., Shikin~V., Sov. Phys. JETP, 1975, \textbf{41}, No. 4, 710--714.  
	
\bibitem{prl1979FisherHP} Fisher~D.S., Halperin~B.I., Platzman~P.M., Phys. Rev. Lett., 1979, \textbf{42}, 798, \doi{10.1103/PhysRevLett.42.798}.
	
\bibitem{pr1934Wigner} Wigner~E., Phys. Rev., 1934, \textbf{46}, 1002--1011, \doi{10.1103/PhysRev.46.1002}.
	
\bibitem{ltp1998PPS} Peletminsky~A.S., Peletminsky~S.V., Slyusarenko~Yu.V., Low Temp. Phys., 1999,  \textbf{25}, No. 5, 303--313,
    \doi{10.1063/1.593743}.
	
\bibitem{prl1979GrimesA} Grimes~C.C., Adams~G., Phys. Rev. Lett., 1979,  \textbf{42}, No. 12, 795--798, \doi{10.1103/PhysRevLett.42.795}.

	
\bibitem{pla1979LeidererW} Leiderer~P., Wanner~M., Phys. Lett. A, 1979, \textbf{73}, No. 3, 189--192, \doi{10.1016/0375-9601(79)90704-7}.

\bibitem{ss1982LeidererES} Leiderer~P., Ebner~W., Shikin~V.B., Surf. Sci., 1982, \textbf{113}, 405--411, \doi{10.1016/0039-6028(82)90623-9}.

\bibitem{pla1980EbnerL} Ebner~W.,  Leiderer~P., Phys. Lett. A, 1980, \textbf{80}, No. 4, 277--280, \doi{10.1016/0375-9601(80)90021-3}.

\bibitem{jetpl1979TroyanovskiiVH} Troyanovskii~A.M., Volodin~A.P., Khaikin~M.S., JETP Lett., 1979,  \textbf{29}, No. 1, 59--62.
	
\bibitem{ss1984Kajita} Kajita~K., Surf. Sci., 1984, \textbf{142}, 86--95, \doi{10.1016/0039-6028(84)90290-5}.
	
\bibitem{prb1982BishopDT} Bishop~D.J., Dynes~R.C., Tsui~D.C., Phys. Rev. B, 1982, \textbf{26}, No. 2, 773--779, \doi{10.1103/PhysRevB.26.773}.
	
\bibitem{book1997Andrei} Andrei~E.Y., 2D Electron Systems on Helium and Other Cryogenic Substrates, Kluwer, Dordrecht, 1997.
	
\bibitem{book2003MonarkhaK} Monarkha~Yu., Kono~K., Two-Dimensional Coulomb Liquids and Solids, Springer-Verlag, Berlin, 2003. 
	
\bibitem{ltp1982MonarkhaS} Monarkha~Yu.P., Shikin~V.B., Fiz. Nizk. Temp., 1982, \textbf{8}, No. 6, 563--601 (in Russian). 
	
\bibitem{ltp2012MonarkhaS} Monarkha~Yu.P., Syvokon~V.E., Low Temp. Phys., 2012, \textbf{38}, No. 12, 1067--1095,
	\doi{10.1063/1.4770504}.
	
\bibitem{ufn2011Shikin} Shikin~V.B., Usp. Fiz. Nauk, 2011, \textbf{181}, No. 12, 1241--1264 (in Russian),
	\doi{10.3367/UFNr.0181.201112a.1241}.
	
\bibitem{prl1969ColeC} Cole~M.W., Cohen~M.H., Phys. Rev. Lett., 1969, \textbf{23}, No. 21, 1238--1241, \doi{10.1103/PhysRevLett.23.1238}.
	
\bibitem{jmp2012LytvynenkoSK} Lytvtnenko~D.M., Slyusarenko~Yu.V., Kirdin~A.I., J. Math. Phys., 2012, \textbf{53}, 103302, \doi{10.1063/1.4753978}.
	
\bibitem{past2012SlyusarenkoSK} Slyusarenko~Yu.V., Lytvynenko~D.M., Kirdin~A.I., Prob. At. Sci. Technol., 2012, No.~1, 288--291. 
	
\bibitem{cmp2009LytvynenkoS} Lytvynenko~D.M., Slyusarenko~Yu.V., Condens. Matter Phys., 2009, \textbf{12}, No. 1, 19--34, \doi{10.5488/CMP.12.1.19}
	
\bibitem{jps2015SlyusarenkoL} Slyusarenko~Yu.V., Lytvynenko~D.M., J. Phys. Stud., 2015, \textbf{19}, No. 3, 3601. 
	
\bibitem{jpa2017LytvynenkoS} Lytvynenko D.M., Slyusarenko Yu.V., J. Phys. A: Math. Theor., 2017, \textbf{50}, 315202,\\ \doi{10.1088/1751-8121/aa76ab}.
	
\bibitem{pre1998LevZ} Lev~B.I., Zhugaevych~A.Ya., Phys. Rev. E, 1998, \textbf{57}, No. 6, 6460--6469, \doi{10.1103/PhysRevE.57.6460}.
	
	
\bibitem{pre2011LevZ} Lev~B.I., Zagorodny~A.G., Phys. Rev. E, 2011, \textbf{84}, 061115, \doi{10.1103/PhysRevE.84.061115}.
	
	
\bibitem{ujp2015LevOTZ} Lev~B.I., Ostroukh~V.P., Tymchyshyn~V.B., Zagorodny~A.G., Ukr. J. Phys., 2015, \textbf{60}, 247--252, \\ \doi{10.15407/ujpe60.03.0247}.
	
\bibitem{epjb2014LevOTZ} Lev~B.I., Ostroukh~V.P., Tymchyshyn~V.B., Zagorodny~A.G., Eur. Phys. J. B, 2014, \textbf{87}, 253,\\ \doi{10.1140/epjb/e2014-50303-2}.

\bibitem{book1989ShikinM} Shikin V.B., Monarkha Yu.P., 2D Charged Systems in Helium, Nauka, Moscow, 1989, (in Russian).

\bibitem{book1970Landau5} Landau L.D., Lifshitz E.M., Statistical Physics, 2nd Edn., Part 1, Vol. 5, Pergamon, Oxford, 1970.

\bibitem{prl1981GianettaI} Giannetta R.W., Ikezi H., Phys. Rev. Lett., 1981, \textbf{47}, No. 12, 849--852, \doi{10.1103/PhysRevLett.47.849}.

\bibitem{book1977Landau3} Landau L.D., Lifshitz E.M., Quantum Mechanics, 3rd Edn.,  Vol. 3, Pergamon, Oxford, 1977.

\end{thebibliography}
\end{document}